\documentclass[doublecol]{epl2} 
\usepackage{amsmath,amsfonts,amssymb,mathrsfs,graphicx,color,longtable}
\usepackage{bm}
\usepackage{array}
\usepackage{color}
\usepackage{enumitem}
\usepackage{ulem}
\usepackage[utf8]{inputenc}
\usepackage{float}  
\usepackage{multirow,bigdelim}
\usepackage{blindtext}
\usepackage{rotating}

\usepackage[spaces,hyphens]{url}
\usepackage[colorlinks,allcolors=blue]{hyperref} %% optional

\usepackage[TABBOTCAP,tight]{subfigure}
\newcommand{\vet}[1]{\ensuremath{\hskip-1pt\vec{\hskip1pt#1}}}

\title{A view of Coherent Elastic Neutrino-Nucleus Scattering}
\shorttitle{A view of CE$\nu$NS} %Insert here a short version of the title if it exceeds 70 characters

\author{M. Cadeddu\inst{1} \and F. Dordei\inst{1} \and C. Giunti\inst{2}}
\shortauthor{M. Cadeddu \etal}

\institute{                    
\inst{1} Istituto Nazionale di Fisica Nucleare (INFN), Sezione di Cagliari,
Complesso Universitario di Monserrato - S.P. per Sestu Km 0.700,
09042 Monserrato (Cagliari), Italy
\\
\inst{2} Istituto Nazionale di Fisica Nucleare (INFN), Sezione di Torino,
Via P. Giuria 1, 10125 Torino, Italy
}

\abstract{We review the physics of coherent elastic neutrino-nucleus scattering
and the results and perspectives for
the measurements
of the radius of the neutron distribution of the nucleus,
of the weak mixing angle,
and
of new neutrino interactions due to physics beyond the Standard Model.}

\begin{document}

\maketitle

\section{Introduction}
\label{sec:intro}

Coherent elastic neutrino-nucleus scattering (CE$\nu$NS)
is a peculiar interaction process which was predicted theoretically in 1973~\cite{Freedman:1973yd}
(see also~\cite{Freedman:1977xn,Drukier:1984vhf}),
but it was observed experimentally for the first time only in 2017
in the COHERENT experiment~\cite{COHERENT:2017ipa},
with neutrinos produced by the Spallation Neutron Source (SNS) at the Oak Ridge National Laboratory.
The cross-section of CE$\nu$NS is larger than the cross-sections of other neutrino interactions at the same neutrino energy.
Thus, one may wonder why it took 44 years to observe CE$\nu$NS after its theoretical prediction.
The reason is that the only observable signature of CE$\nu$NS is the recoil of the nucleus with an extremely small kinetic energy which is very difficult to observe.
The recent development of new very sensitive detectors allowed this magnificent achievement.

The main characteristic of CE$\nu$NS is that the interactions of the neutrino with the nucleons must be coherent,
such that the nucleus recoils as a whole,
without any change in its internal structure.
In this case, the predicted Standard Model (SM) weak-interaction cross-section is approximately
proportional to the square of the number of neutrons $N$ in the nucleus.
Therefore, the probability of CE$\nu$NS interactions is approximately
$N$ times larger than the sum of the probabilities of neutrino
interaction with every single nucleon in the same nucleus,
which is only approximately proportional to $N$.
The reason why the number of neutrons $N$ in the nucleus is more relevant
than the number of protons
is that, as explained later,
the cross-section of neutrino-proton interaction
is much smaller than the cross-section of neutrino-neutron interaction.

CE$\nu$NS is a process that is not only interesting per se.
It gives useful information on the nuclear structure, in particular on the neutron density distribution in the nucleus,
which surprisingly is still unknown for most of the nuclei.
It gives also information on the precise value of the
neutrino neutral-current interaction at low energy,
which is quantified by the so-called ``weak mixing angle''.
Moreover, CE$\nu$NS is sensitive to effects of physics Beyond the Standard Model (BSM)
such as
neutrino electromagnetic interactions,
neutrino interactions mediated by new neutral BSM bosons,
and, in general, neutrino neutral-current nonstandard interactions (NSI).

So far, all the results obtained from CE$\nu$NS experiments are compatible with the SM prediction.
After the first observation by the COHERENT Collaboration of CE$\nu$NS in 2017 using a cesium-iodide (CsI) detector~\cite{COHERENT:2017ipa},
a larger dataset of scattering events have been collected with the same 
experiment~\cite{COHERENT:2021xmm}
and the same process has also been observed with a liquid argon detector~\cite{COHERENT:2020iec}.
Recently, the experimental collaboration working at the Dresden-II experiment
reported the observation of CE$\nu$NS
produced by electron antineutrinos from a nuclear power reactor~\cite{Colaresi:2022obx}.
Other experiments
(CONUS~\cite{CONUS:2020skt} and CONNIE~\cite{CONNIE:2021ggh})
have been able to obtain upper limits on CE$\nu$NS,
which are compatible with the SM prediction, and are still working towards their first observation.
Several other CE$\nu$NS experiments are planned or under construction:
new COHERENT detectors~\cite{Akimov:2022oyb},
MINER~\cite{MINER:2016igy},
NEON~\cite{NEON:2022hbk},
NUCLEUS~\cite{NUCLEUS:2019igx},
RED-100~\cite{Akimov:2022xvr},
Ricochet~\cite{Colas:2021pxr},
a project at the European Spallation Source~\cite{Baxter:2019mcx},
and others~\cite{Abdullah:2022zue}.
Therefore, we expect many new experimental results in the next years,
which will bring very interesting information on
nuclear physics, weak neutral-current neutrino interactions and
BSM physics.

\section{CE$\mathbf{\nu}$NS in the Standard Model}
\label{sec:SM}

\begin{figure}
\centering
\subfigure[]{\label{fig:CEvNS-Feynman}
\includegraphics[width=0.23\textwidth]{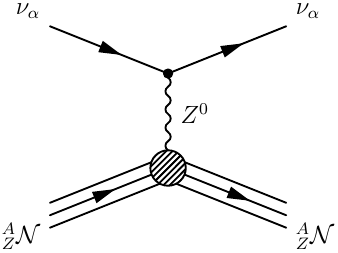}}
\subfigure[]{\label{fig:CEvNS-BSM}
\includegraphics*[width=0.23\textwidth]{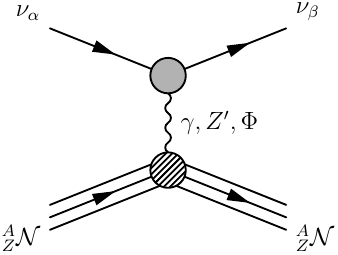}
}
\caption{ (a) Feynman diagram of CE$\nu$NS in the Standard Model.
$\nu_{\alpha}$ represents a neutrino with flavor $\alpha=e,\,\mu,\,\text{or}\,\tau$,
${}^{A}_{Z}\mathcal{N}$ denotes a nucleus
with $A$ nucleons ($Z$ protons and $N=A-Z$ neutrons),
and $Z^{0}$ is the Standard Model neutral vector boson
which mediates neutral-current weak interactions.
(b)
Feynman diagram of contributions to CE$\nu$NS of BSM physics in which the interaction is mediated by a photon $\gamma$,
or a new massive vector boson $Z'$,
or scalar boson $\Phi$.}
\end{figure}

CE$\nu$NS is the neutral-current process
\begin{equation}
\nu_{\alpha} + {}^{A}_{Z}\mathcal{N}
\to
\nu_{\alpha} + {}^{A}_{Z}\mathcal{N}
,
\label{eq:CEvNSproc}
\end{equation}
where $\nu_{\alpha}$ represents a neutrino with flavor $\alpha=e,\,\mu,\,\text{or}\,\tau$,
and ${}^{A}_{Z}\mathcal{N}$ denotes a nucleus
with $A$ nucleons, of which $Z$ are protons and $N=A-Z$ are neutrons.
Neutral-current processes are characterized by the absence of an exchange of electric charge between the interacting particles.
Hence, they are mediated by the exchange of a neutral boson.
In the SM the mediator is the $Z^{0}$ neutral boson,
as illustrated in Fig.~\ref{fig:CEvNS-Feynman}.

%\begin{figure}
%\begin{center}
%\includegraphics[width=0.5\linewidth]{figures/CEvNS-Feynman.pdf}
%%\onefigure{figures/CEvNS-Feynman.pdf}
%\caption{Feynman diagram of CE$\nu$NS in the Standard Model. $\nu_{\alpha}$ represents a neutrino with flavor $\alpha=e,\,\mu,\,\text{or}\,\tau$, ${}^{A}_{Z}\mathcal{N}$ denotes a nucleus with $A$ nucleons ($Z$ protons and $N=A-Z$ neutrons), and $Z^{0}$ is the Standard Model neutral vector boson which mediates neutral-current weak interactions.}
%\label{fig:CEvNS-Feynman}
%\end{center}
%\end{figure}

%\begin{figure}
%\begin{center}
%\includegraphics[width=0.5\linewidth]{figures/CEvNS-BSM.pdf}
%%\onefigure{figures/CEvNS-BSM.pdf}
%\caption{Feynman diagram of contributions to CE$\nu$NS of BSM physics in which the interaction is mediated by a photon $\gamma$, or a new massive vector boson $Z'$, or scalar boson $\Phi$.}
%\label{fig:CEvNS-BSM}
%\end{center}
%\end{figure}

\begin{figure*}
\begin{center}
\begin{tabular}{ccc}
\includegraphics[clip, width=0.18\linewidth]{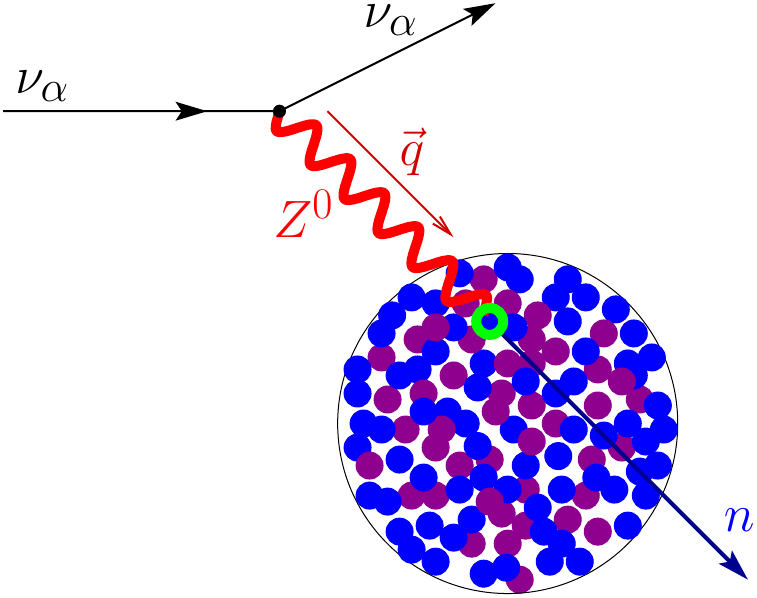}
&
\includegraphics[clip, width=0.18\linewidth]{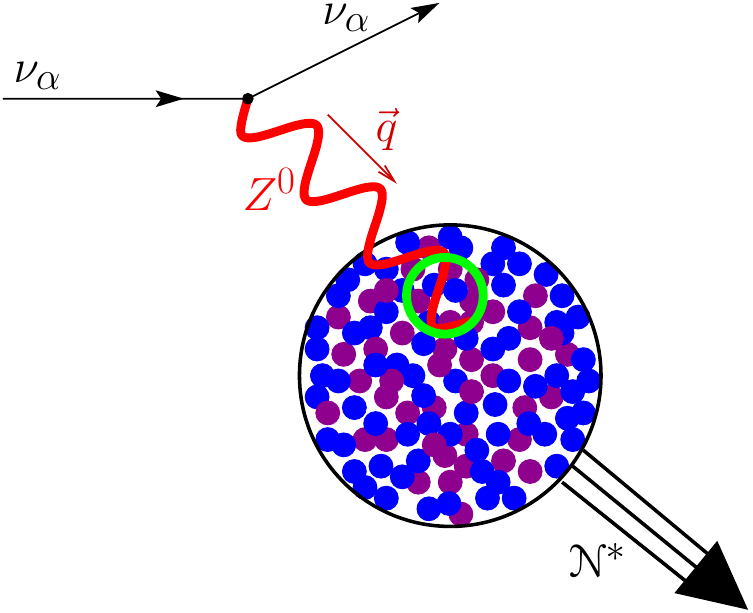}
&
\includegraphics[clip, width=0.18\linewidth]{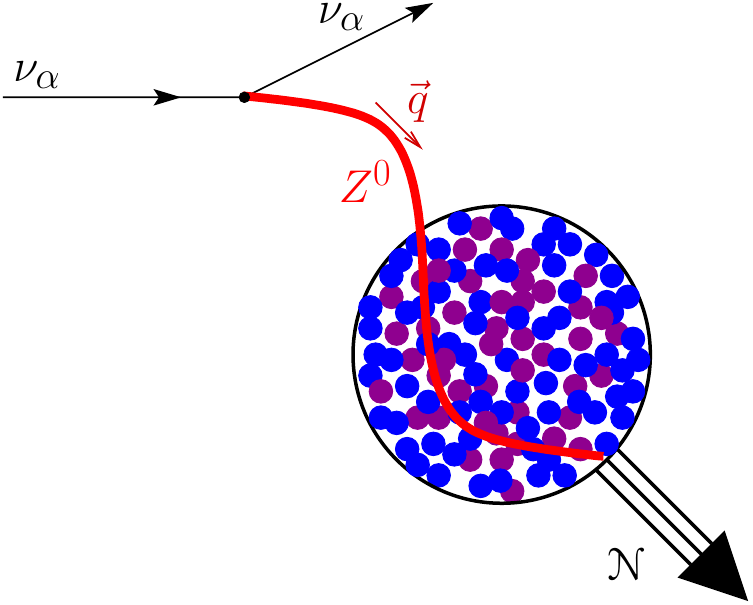}
\\[0.3cm]
Inelastic incoherent
&
Elastic incoherent
&
Elastic coherent (CE$\nu$NS)
\\%[0.3cm]
$\displaystyle
\lambda_{Z^{0}} \ll 2R
$
&
$\displaystyle
\lambda_{Z^{0}} \lesssim 2R
$
&
$\displaystyle
\lambda_{Z^{0}} \gtrsim 2R
$
\end{tabular}
\end{center}
\caption{Illustration of three types of interactions of a neutrino $\nu_{\alpha}$ with a nucleus:
inelastic incoherent scattering when $\lambda_{Z^{0}} \ll 2R$,
elastic incoherent scattering when $\lambda_{Z^{0}} \lesssim 2R$, and
elastic coherent scattering (CE$\nu$NS) when $\lambda_{Z^{0}} \gtrsim 2R$.
Here $R$ is the radius of the nucleus and
$\lambda_{Z^{0}} = h / |\vet{q}|$
is the wavelength of the $Z^{0}$ neutral vector boson,
where $\vet{q}$ is its three-momentum and $h$ is Planck's constant.}
\label{fig:scattering}
\end{figure*}

Figure~\ref{fig:scattering}
illustrates three types of interactions of a neutrino $\nu_{\alpha}$ with a nucleus,
which depend on the wavelength $\lambda_{Z^{0}} = h / |\vet{q}|$ of the $Z^{0}$ neutral vector boson which mediates SM neutral-current weak interactions ($h$ is Planck's constant and $\vet{q}$ is the $Z^{0}$ three-momentum).
When $\lambda_{Z^{0}} \ll 2R$,
where $R$ is the radius of the nucleus,
the $Z^{0}$ has a high probability to interact with a single nucleon in the nucleus
(a neutron in the illustration),
ejecting it.
When $\lambda_{Z^{0}} \lesssim 2R$
the $Z^{0}$ has a high probability to interact with a group of nucleons in the nucleons,
exciting the latter to the state $\mathcal{N}^{*}$.
Elastic coherent neutrino-nucleus scattering can happen only when
$\lambda_{Z^{0}} \gtrsim 2R$.
In this case,
the wavelength of the $Z^{0}$ covers all the nucleus,
allowing a coherent nuclear recoil without any internal change of the nucleus.

For CE$\nu$NS, the absolute value $|\vet{q}|$ of the $Z^{0}$ three-momentum
is related to the measurable value of the kinetic energy of the recoiling nucleus $T$ by
$ |\vet{q}| \simeq \sqrt{ 2 \, M \, T } $,
where $M$ is the mass of the nucleus.
Therefore,
the coherency condition is satisfied only for very small values of $T$,
which are difficult to measure.
For example,
let us consider a heavy nucleus with $A \approx 100$ nucleons,
for which the nuclear radius can be approximately estimated
with the well-known semi-empirical formula
$R \approx 1.2 \, A^{1/3} \, \text{fm} \approx 5 \, \text{fm}$.
Therefore, CE$\nu$NS can happen for
$|\vet{q}| \lesssim 40 \, \text{MeV}$,
which implies
$T \lesssim 10 \, \text{keV}$.
This is a very low value of the recoil energy of the nucleus,
which could not be measured before the recent technical developments
of very sensitive detectors employed in the search of dark matter in the form of weakly interacting massive particles (WIMPs). Indeed, WIMPs with masses of $\mathcal{O}(10\,\mathrm{GeV/}c^2)$ induce nuclear recoils in the keV range which are very similar to those produced by CE$\nu$NS processes.

Moreover,
since the average momentum transfer depends on the neutrino energy,
it is only possible to observe CE$\nu$NS generated by low-energy neutrinos
($ E_{\nu} \lesssim 30 \, \text{MeV} $).
There are various natural sources of low-energy neutrinos:
solar neutrinos,
geoneutrinos,
supernova neutrinos,
and the low-energy tail of atmospheric neutrinos~\cite{Abdullah:2022zue}.
Several experiments are sensitive to CE$\nu$NS generated by these natural low-energy
neutrino fluxes (especially experiments on the search of nuclear recoils
produced by dark matter~\cite{AristizabalSierra:2019ykk}),
but so far these neutrino fluxes have been observed only through other interactions.

CE$\nu$NS has been observed only in two laboratory experiments:
\begin{enumerate}

\item
The COHERENT experiment~\cite{COHERENT:2017ipa,COHERENT:2020iec,COHERENT:2021xmm},
with neutrinos produced by the Oak Ridge SNS.
In this experiment, the neutrinos are generated by the decay at rest of positive pions ($\pi^{+}$) and muons ($\mu^{+}$).
The positive pions are created by the interaction of a pulsed beam of protons with an energy of about 1~GeV with a mercury target
(the produced negative pions are rapidly absorbed by the positive nuclei of the target before decay).
They stop in the target
and decay at rest with a lifetime of about 26~ns with the process
$\pi^{+} \to \mu^{+} + \nu_{\mu}$,
generating a prompt flux of muon neutrinos ($\nu_{\mu}$).
The produced positive muons also stop in the target
and decay at rest with a longer lifetime of about 2.2~$\mu s$ with the process
$\mu^{+} \to e^{+} + \bar\nu_{\mu} + \nu_{e}$,
generating delayed fluxes of muon antineutrinos ($\bar\nu_{\mu}$)
and electron neutrinos ($\nu_{e}$).
Since the energies of these neutrino fluxes are below about 53 MeV,
they generate observable rates of CE$\nu$NS in the COHERENT
CsI~\cite{COHERENT:2017ipa,COHERENT:2021xmm}
and LAr~\cite{COHERENT:2020iec}
detectors, where the total neutrino flux is about
$4.3 \times 10^{7} \, \nu \, \text{cm}^{-2} \, s^{-1}$.
It is remarkable that the detectors are very small in comparison with
typical neutrino detectors, with mass scales of the order of multiples of tons.
The CsI detector weighs only 14.6 kg
and the LAr detector weighs 29 kg.
The detection of CE$\nu$NS with such small detectors is possible
because of the large CE$\nu$NS cross-section.

\item
The Dresden-II experiment~\cite{Colaresi:2022obx}
with electron antineutrinos ($\bar\nu_{e}$)
generated by the Dresden-II nuclear power reactor
of the Dresden Generating Station
located near Morris, Illinois, USA.
The flux of electron antineutrinos at the very sensitive germanium detector
is huge,
about $4.8 \times 10^{13} \, \bar\nu_{e} \, \text{cm}^{-2} \, s^{-1}$,
with energy below about 10 MeV.
It is well suited to generate an observable rate of CE$\nu$NS in the
extraordinarily small germanium detector weighing only 3 kg. We note that there is, however, some criticism of this result related to the anomalously large quenching factor, the reactor-related background  and the incompatibility with the recent CONUS results~\cite{conusmag} which must be clarified~\cite{expsummary}.

\end{enumerate}

The measurable CE$\nu$NS differential cross-section
of a neutrino with a spin-zero nucleus $\mathcal{N}$ is
\begin{equation}
\dfrac{d\sigma_{\nu\mathcal{N}}}{d T}(E_{\nu},T)
=
\dfrac{ G_{\text{F}}^2 M }{ \pi }
\left(
1 - \dfrac{ M T }{ 2 E_{\nu}^2 }
\right)
\left[ Q_{W}^{\mathcal{N}}(|\vet{q}|) \right]^2
,
\label{eq:cs}
\end{equation}
where
$E_{\nu}$ is the neutrino energy, and
$G_{\text{F}}$ is the Fermi constant.
The interaction is characterized by the so-called ``weak charge of the nucleus''
$Q_{W}^{\mathcal{N}}(|\vet{q}|)$,
a function of the three-momentum $\vet{q}$
transferred from the neutrino to the nucleus
that depends on the interaction type.
In the SM
\begin{equation}
Q_{W}^{\mathcal{N}}(|\vet{q}|)
=
g_{V}^{n}
\,
N
\,
F_{N}^{\mathcal{N}}(|\vet{q}|)
+
g_{V}^{p}
\,
Z
\,
F_{Z}^{\mathcal{N}}(|\vet{q}|)
.
\label{eq:SMQW}
\end{equation}
The functions
$F_{N}^{\mathcal{N}}(|\vet{q}|)$
and
$F_{Z}^{\mathcal{N}}(|\vet{q}|)$
are, respectively, the neutron and proton form factors of the nucleus $\mathcal{N}$,
which characterize the amount of coherency of the interaction
and depend on the neutron and proton distributions in the nucleus.
The coefficients
$g_{V}^{n}$ and $g_{V}^{p}$
quantify the weak neutral-current interactions of neutrons and protons,
respectively.
In the SM they are approximately given by\footnote{The more accurate numerical values which incorporate the so-called radiative corrections can be found in \cite{AtzoriCorona:2023ktl}.}
\begin{equation}
g_{V}^{n} \simeq - \frac{1}{2}
\quad
\text{and}
\quad
g_{V}^{p}
\simeq
\dfrac{1}{2} - 2 \sin^2\!\vartheta_{W}
\simeq
0.022
,
\label{eq:gV}
\end{equation}
where $\vartheta_{W}$ is the weak mixing angle,
with $\sin^2\!\vartheta_{W} \simeq 0.239$.
Therefore,
the neutron contribution is much larger than the proton contribution
and the CE$\nu$NS cross section is approximately proportional to
$N^2$.

\begin{figure}
\begin{center}
\includegraphics[width=0.8\linewidth]{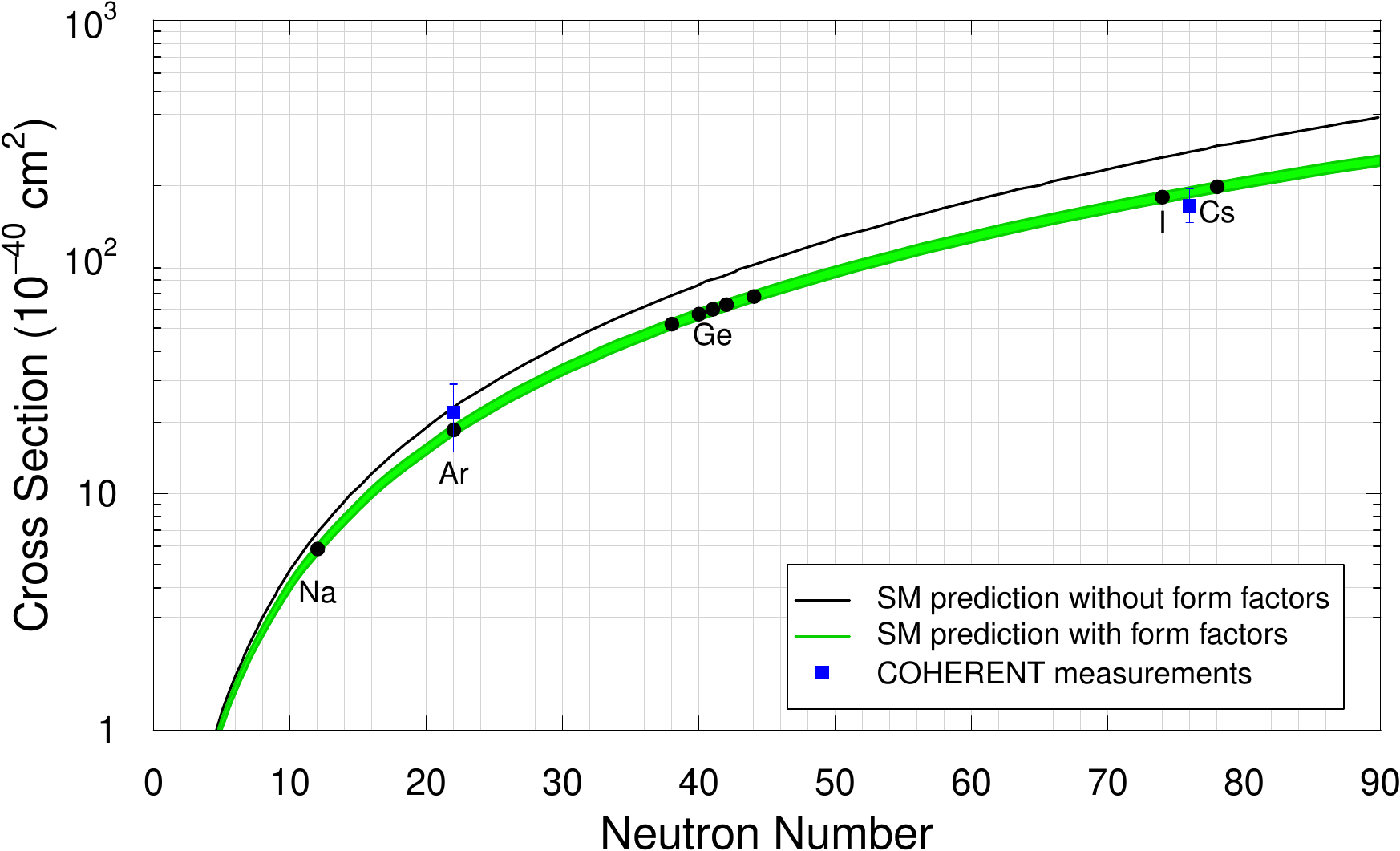}
%\onefigure[width=\linewidth]{figures/COHERENT-200310630-Fig_11.pdf}
\caption{Approximate increase of the total CE$\nu$NS cross-section with the square
of the number of neutrons $N$ in the nucleus $\mathcal{N}$
(adapted from the Fig. 11 in Ref.~\cite{COHERENT:2020iec}).
The black curve shows the increase with $N^2$
in the case of full coherency (i.e.
$F_{N}^{\mathcal{N}}(|\vet{q}|) = 1$
in Eq.~\eqref{eq:SMQW}),
whereas the green curve shows the expected increase which is slightly slower
because of a small amount of incoherency quantified by a realistic form factor
$F_{N}^{\mathcal{N}}(|\vet{q}|)$.
The black dots correspond to the predictions for four selected nuclei.
The blue dots with error bars show the measurements of the COHERENT experiment.}
\label{fig:COHERENTcs}
\end{center}
\end{figure}

\section{Nuclear Physics}
\label{sec:Nuclear}

The approximate increase of the total CE$\nu$NS cross-section with the square
of the number of neutrons in the nucleus
is shown by the black curve in Fig.~\ref{fig:COHERENTcs}.
However, one can see that the result of the COHERENT
measurement of the total CE$\nu$NS cross-section with the CsI detector
lies below the black line and is compatible with the green line
which takes into account a small amount of incoherency quantified by a realistic form factor
$F_{N}^{\mathcal{N}}(|\vet{q}|)$.
Indeed, the COHERENT CsI data show a 6$\sigma$ evidence of the nuclear structure suppression of the full coherence~\cite{AtzoriCorona:2023ktl}.

The dependence of the cross-section on the neutron form factor
$F_{N}^{\mathcal{N}}(|\vet{q}|)$
is a powerful tool for obtaining information on the neutron distribution
in the nucleus,
which is not well-known.
Indeed, while the proton distribution of most nuclei
is known from electromagnetic measurements~\cite{Angeli:2013epw}, the nuclear neutron distribution
can be probed only through measurements which employ the strong or weak
forces.
The interpretation of the results of experiments with hadron probes
which explore the nuclear neutron distribution through the strong force
is difficult, since the effects of strong-force interactions
cannot be calculated with sufficient approximation
and the interpretation can be done only by assuming a strong-interaction model
with all its limitations~\cite{Thiel:2019tkm}.
On the other hand,
the effects of the weak neutral-current interactions,
embodied by $Q_{W}^{\mathcal{N}}(|\vet{q}|)$ in Eq.~\eqref{eq:SMQW},
are known with good approximation.
Therefore,
weak neutral-current processes as CE$\nu$NS are ideal for
probing the nuclear neutron distribution.
Besides CE$\nu$NS, the nuclear neutron distribution has been probed
with neutral-current weak interactions
through parity-violating electron scattering for only two nuclei:
$^{208}\text{Pb}$
in the PREX experiment~\cite{PREX:2021umo}
and
${^{48}\textrm{Ca}}$
in the CREX experiment~\cite{CREX:2022kgg}.
The quantity which in practice is measured is the radius $R_{n}$ of the
neutron distribution in the nucleus.
Since in heavy nuclei there are more neutrons than protons,
all nuclear models predict that $R_{n}$ should be larger than
the radius $R_{p}$ of the
proton distribution in the same nucleus.
The excess
$ \Delta R_{np} = R_{n} - R_{p} $
is called ``neutron skin''.
It is a quantity of great interest for nuclear physics and astrophysics,
because it is the result of the competition
between  the Coulomb repulsion between the protons,
the surface  tension,
that decreases when the excess neutrons are pushed to the surface,
and the symmetry energy~\cite{Baldo:2016jhp}.
It gives information on the nuclear equation of state,
which determines the size of neutron stars~\cite{Horowitz:2000xj}.

\begin{figure}
\begin{center}
\includegraphics[width=0.75\linewidth]{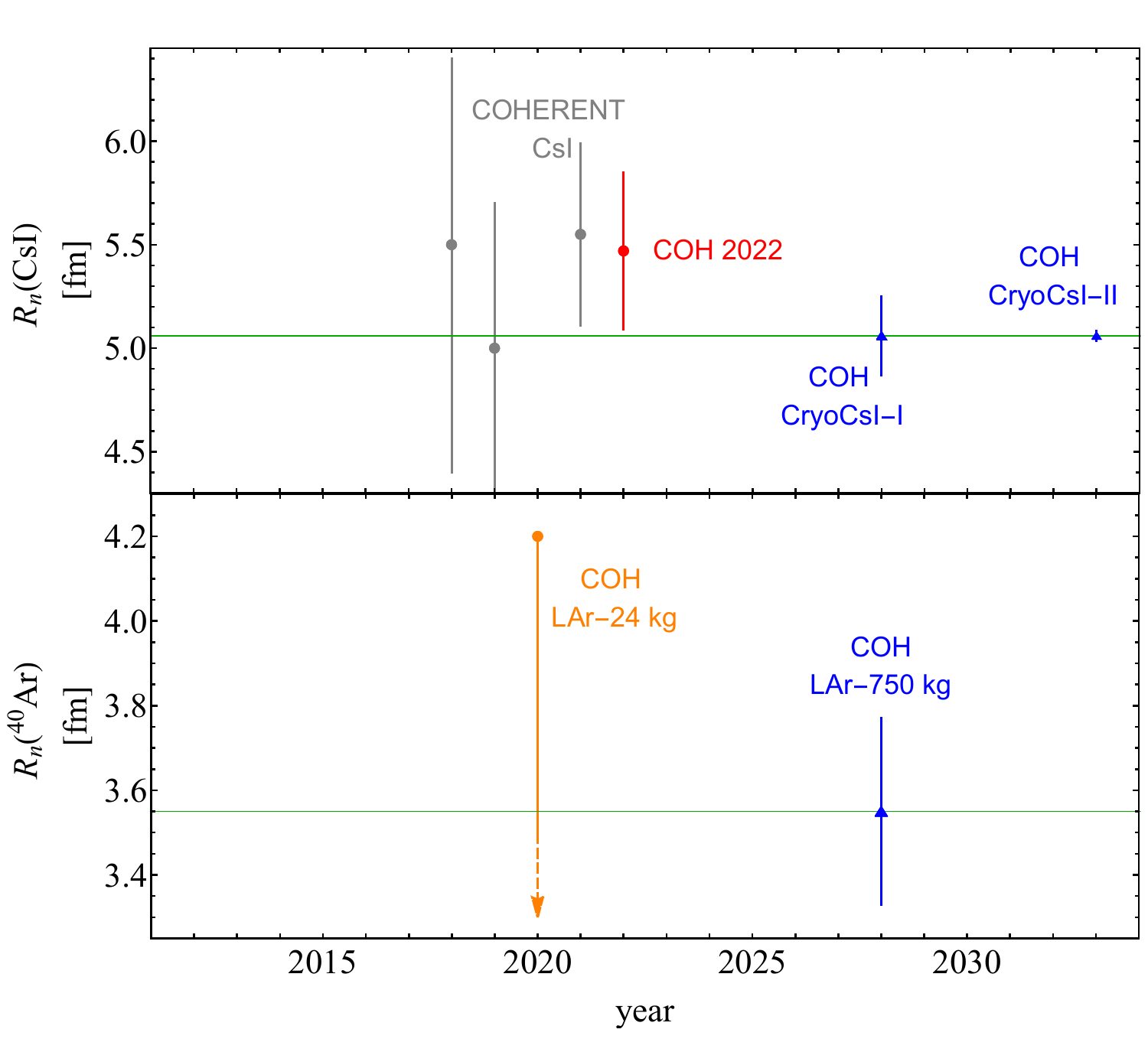}
%\onefigure[width=\linewidth]{figures/plotRiassuntoRn.pdf}
\caption{Current status and future projections for neutron distribution radii of different nuclei measured via CE$\nu$NS. The upper and lower plots show,
respectively,
the current and foreseen measurements of $R_{n}^{\text{CsI}}$ from CE$\nu$NS with CsI detectors
\cite{Cadeddu:2017etk,Cadeddu:2019eta,Cadeddu:2021ijh,AtzoriCorona:2023ktl}
and those of $R_{n}^{\text{Ar}}$ with Ar detectors
\cite{Cadeddu:2020lky,Akimov:2022oyb}.}
\label{fig:plotRiassuntoRn}
\end{center}
\end{figure}

For CE$\nu$NS,
the neutron form factor
$F_{N}^{\mathcal{N}}(|\vet{q}|)$, being the Fourier transform of the neutron distribution in
the nucleus, 
depends on $R_{n}$.
The CE$\nu$NS measurements of the COHERENT collaboration
with the CsI detector~\cite{COHERENT:2017ipa,COHERENT:2021xmm}
give information on the average radius $R_{n}^{\text{CsI}}$
of the neutron distributions in
$^{133}\text{Cs}$ and $^{127}\text{I}$
\cite{Cadeddu:2017etk,Cadeddu:2018dux,Cadeddu:2019eta,Papoulias:2019lfi,Khan:2019cvi,Huang:2019ene,Coloma:2020nhf,AtzoriCorona:2023ktl}:
$R_{n}^{\text{CsI}} = 5.47 \pm 0.38 \, \text{fm}$~\cite{AtzoriCorona:2023ktl}.
This value
is in agreement, within the uncertainty,
with the recent Nuclear Shell Model (NSM) prediction
$R_{n}^{\text{CsI}}(\text{NSM}) \simeq 5.06 \, \text{fm}$~\cite{Hoferichter:2020osn,Cadeddu:2021ijh}.
The CE$\nu$NS measurement of the COHERENT collaboration
with the LAr detector~\cite{COHERENT:2020iec}
has larger uncertainties and leads only to an upper bound for
the radius $R_{n}^{\text{Ar}}$ of the neutron distributions in
$^{40}\text{Ar}$~\cite{Miranda:2020tif,Cadeddu:2020lky}:
$R_{n}^{\text{Ar}} < 4.2 \, \text{fm}$~\cite{Cadeddu:2020lky}
at $1\sigma$ (i.e. 68\% confidence level).
This bound is compatible with the recent NSM prediction
$R_{n}^{\text{Ar}} \simeq 3.6 \, \text{fm}$~\cite{Hoferichter:2020osn}.
Further improvements in the determination of
$R_{n}^{\text{CsI}}$ and $R_{n}^{\text{Ar}}$
are expected when new measurements of the COHERENT collaboration
will be available~\cite{Akimov:2022oyb},
as illustrated in Fig.~\ref{fig:plotRiassuntoRn}.

An important improvement in the determination of $R_{n}^{\text{CsI}}$
can be achieved by combining the COHERENT CsI results
with those of Atomic Parity Violation (APV) experiments
with Cs atoms~\cite{Wood:1997zq,Guena:2004sq},
which are sensitive to
$R_{n}^{\text{Cs}}$
\cite{Cadeddu:2018izq,Cadeddu:2021ijh,AtzoriCorona:2023ktl}.
This method permits to disentangle the measurements of the
radii of the neutron distributions of
$^{133}\text{Cs}$ and $^{127}\text{I}$:
$R_{n}^{\text{Cs}} = 5.32^{+0.30}_{-0.23} \, \text{fm}$
and
$R_{n}^{\text{I}} = 5.30^{+0.3}_{-0.6} \, \text{fm}$~\cite{AtzoriCorona:2023ktl}.
Obviously $R_{n}^{\text{Cs}}$ is better determined than $R_{n}^{\text{I}}$.
Both are compatible with the NSM predictions
$R_{n}^{\text{Cs}}(\text{NSM}) \simeq 5.09 \, \text{fm}$
and
$R_{n}^{\text{I}}(\text{NSM}) \simeq 5.03 \, \text{fm}$
\cite{Hoferichter:2020osn,Cadeddu:2021ijh}.

Note that the results of reactor neutrino CE$\nu$NS experiments,
as those of the Dresden-II experiment~\cite{Colaresi:2022obx},
do not give information on the neutron distribution of the target nucleus,
because the neutrino energy is low,
of a few MeV,
and there is an observable CE$\nu$NS signal only for very small momentum transfers.
Therefore,
the interaction is fully coherent and the form factors
in Eq.~\eqref{eq:SMQW}
are practically equal to one.
On the other hand,
in these experiments there is the advantage that
the extraction from the data of other physical quantities
does not depend on the poorly known
neutron distribution of the nucleus.

\section{Electroweak Physics}
\label{sec:Electroweak}
\begin{figure}
\begin{center}
\includegraphics[width=0.75\linewidth]{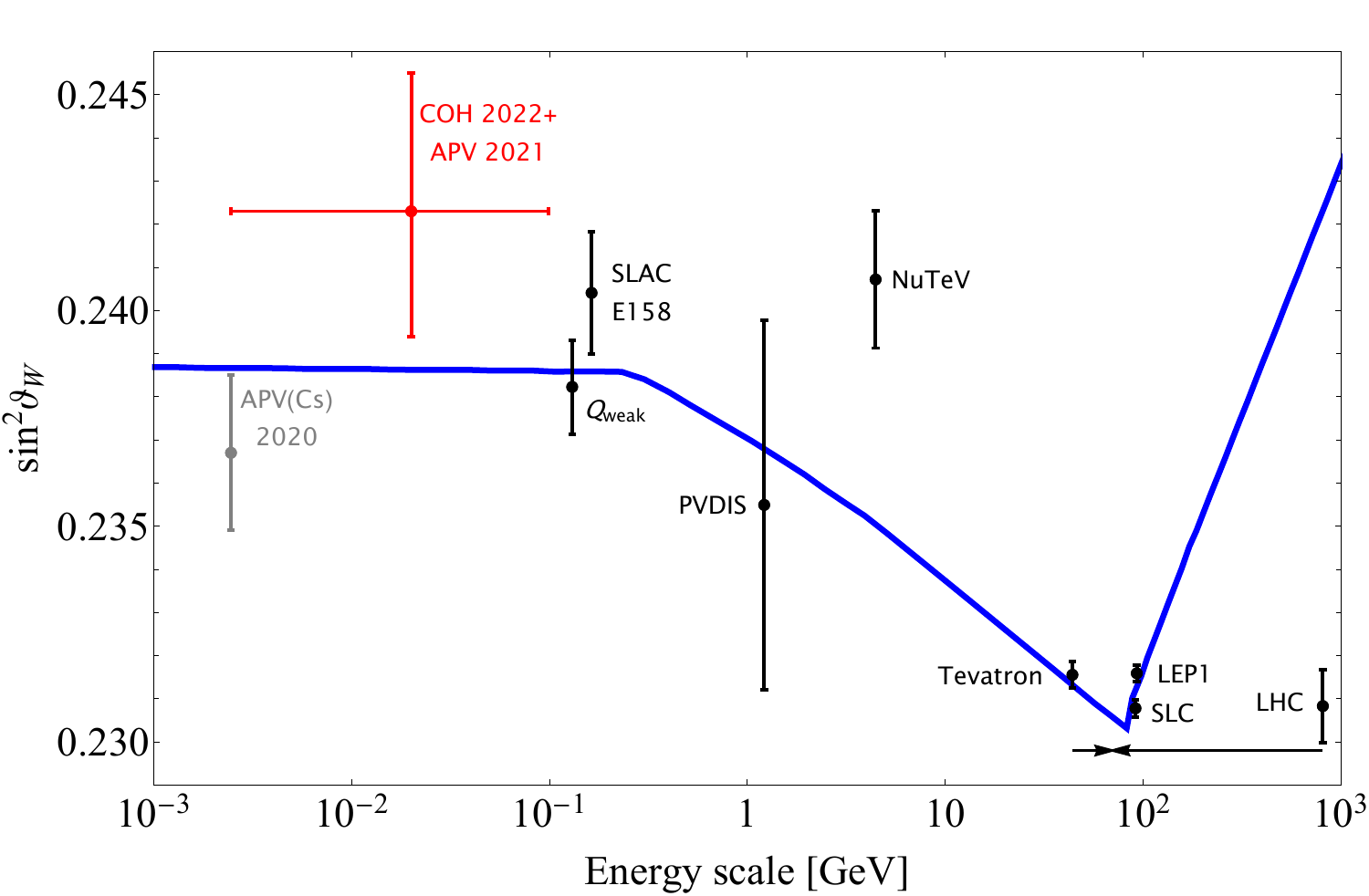}
%\onefigure[width=\linewidth]{figures/Runnings2w_Combinati.pdf}
\caption{Variation of $\sin^2\!\vartheta_{W}$ with the energy scale.
The SM prediction is shown by the blue solid curve,
together with experimental determinations in black~\cite{ParticleDataGroup:2018ovx,Wood:1997zq,Dzuba:2012kx,SLACE158:2005uay,PVDIS:2014cmd,NuTeV:2001whx,Qweak:2018tjf}.
The result obtained in~\cite{AtzoriCorona:2023ktl}
from the combination of COHERENT CE$\nu$NS data and APV data is shown in red.}
\label{fig:Runnings2w_Combinati}
\end{center}
\end{figure}

\begin{figure}
\begin{center}
\includegraphics[width=\linewidth]{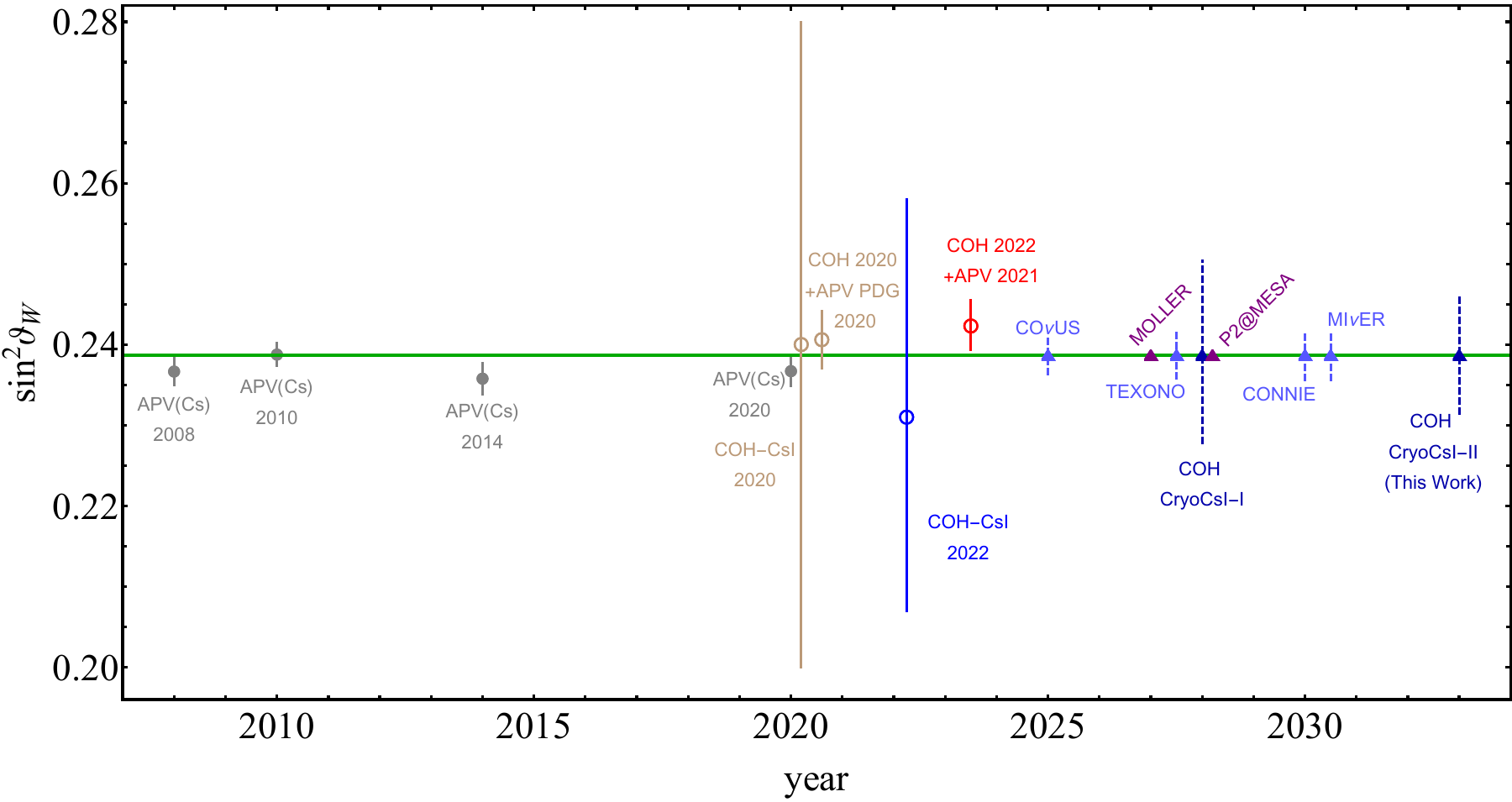}
%\onefigure[width=\linewidth]{figures/pts2wRiassunto.pdf}
\caption{Current status and future projections for low-energy weak mixing angle measurements. The gray points show the measurements from APV on Cs atoms during the years~\cite{ParticleDataGroup:2008zun,ParticleDataGroup:2010dbb,ParticleDataGroup:2014cgo,ParticleDataGroup:2020ssz,ParticleDataGroup:2022pth}. The brown measurements  refer to the COHERENT only and the combination between COHERENT and APV as determined in 2020~\cite{Cadeddu:2021ijh}, while the dark blue and red points refer to the updated measurements reported in~\cite{AtzoriCorona:2023ktl}. The projections for future CE$\nu$NS experiments (CONUS~\cite{Lindner:2016wff}, TEXONO~\cite{TEXONO:2006xds}, CONNIE~\cite{CONNIE:2016nav}, and MINER~\cite{MINER:2016igy}), shown by the light blue triangles and dashed error bars, are taken from~\cite{Canas:2018rng}.
The purple triangles are the projections for the future electron scattering experiment MOLLER~\cite{MOLLER:2014iki} and P2@MESA~\cite{Becker:2018ggl,Dev:2021otb}. The dark blue triangles with dashed error bars are the projections for the future CryoCsI-I and CryoCsI-II determinations~\cite{AtzoriCorona:2023ktl}.}
\label{fig:pts2wRiassunto}
\end{center}
\end{figure}

The CE$\nu$NS cross section depends on the value of $\sin^2\!\vartheta_{W}$
through the proton coupling $g_{V}^{p}$ in Eq.~\eqref{eq:gV}.
Although the proton contribution to the CE$\nu$NS cross section is subdominant
with respect to the neutron one,
it must be taken into account when performing accurate calculations.
In quantum field theory,
the value of $\sin^2\!\vartheta_{W}$ (as well as other quantities) is not a constant,
but depends on the energy scale at which it is measured,
as illustrated in Fig.~\ref{fig:Runnings2w_Combinati}.
This figure shows the SM prediction for $\sin^2\!\vartheta_{W}$
obtained from high-precision measurements
at a scale of about 100 GeV
in experiments at the Tevatron, LEP, SLC and LHC high-energy colliders~\cite{ParticleDataGroup:2018ovx}.
The figure shows also that the low-energy value of $\sin^2\!\vartheta_{W}$
has been probed in several experiments with various results
which have different degrees of compatibility with the SM prediction.
The main interest in measuring the low-energy value of $\sin^2\!\vartheta_{W}$
lies in the possibility of discovering a deviation from the SM
due to new physics contributions~\cite{Safronova:2017xyt,Cadeddu:2021dqx},
such as the mediation of the interaction by dark $Z$ bosons
\cite{Coloma:2022avw,DeRomeri:2022twg,AtzoriCorona:2022moj,Young:2007zs,Cadeddu:2021dqx}.
The determination of the low-energy value of the weak mixing angle
has been discussed in several studies
\cite{Cadeddu:2020lky,Miranda:2020tif,Cadeddu:2021ijh, Cadeddu:2017etk,Papoulias:2019lfi,Papoulias:2017qdn,Cadeddu:2019eta,Papoulias:2019txv,Khan:2019cvi,Dutta:2019eml,AristizabalSierra:2018eqm,Cadeddu:2018izq,Dutta:2019nbn,Abdullah:2018ykz,Ge:2017mcq,Miranda:2021kre,Flores:2020lji,DeRomeri:2022twg,AtzoriCorona:2023ktl},
with the recent updated result
$\sin^2\!\vartheta_{W} = 0.231^{+0.027}_{-0.024}$
\cite{AtzoriCorona:2023ktl},
obtained from the latest COHERENT CsI data~\cite{COHERENT:2021xmm}.
As in the case of the radius of the neutron distribution in the nucleus,
the measurement of $\sin^2\!\vartheta_{W}$
can be improved with a combined analysis of the CE$\nu$NS COHERENT CsI data
and those of APV experiments
with Cs atoms~\cite{Wood:1997zq,Guena:2004sq},
which are also sensitive to the weak mixing angle: $\sin^2\!\vartheta_{W} = 0.2398^{+0.0016}_{-0.0015}$
\cite{AtzoriCorona:2023ktl}\footnote{We note that the APV $\sin^2\!\vartheta_{W}$ determination depends on the  value used for the theoretical parity non-conserving amplitude, for which two different determinations are available~\cite{Sahoo:2021thl,Dzuba:2012kx}. Here, only the value obtained with~\cite{Sahoo:2021thl} has been reported. The particular choice can make a difference as large as 11\%~\cite{AtzoriCorona:2023ktl}.}.
This measurement becomes $\sin^2\!\vartheta_{W} = 0.2423^{+0.0032}_{-0.0029}$~\cite{AtzoriCorona:2023ktl}, as reported in Fig.~\ref{fig:Runnings2w_Combinati}, when the neutron distribution is simultaneously determined. 
Clearly, the current combination is dominated by the APV result,
but future measurements of the COHERENT collaboration~\cite{Akimov:2022oyb}
are expected to reduce significantly the uncertainty
of the CE$\nu$NS determination of the weak mixing angle,
as illustrated in Fig.~\ref{fig:pts2wRiassunto}.
As it can be seen, more $\sin^2\!\vartheta_{W}$ measurements are also expected from other future CE$\nu$NS experiments (CO$\nu$US~\cite{Lindner:2016wff}, TEXONO~\cite{TEXONO:2006xds}, CONNIE~\cite{CONNIE:2016nav}, and MINER~\cite{MINER:2016igy}) that will be really powerful for further constraining such a quantity.

\section{CE$\mathbf{\nu}$NS beyond Standard Model}
\label{sec:BSM}

CE$\nu$NS measurements are powerful probes of physics beyond the SM
which can induce neutral current interactions as those depicted in Fig.~\ref{fig:CEvNS-BSM},
which involve the mediation of a photon $\gamma$,
or a new massive vector boson $Z'$,
or a new massive scalar boson $\Phi$.
Note that in these non-standard interactions
the neutrino flavor can change:
$
\nu_{\alpha} + {}^{A}_{Z}\mathcal{N}
\to
\nu_{\beta} + {}^{A}_{Z}\mathcal{N}
$,
with a final neutrino flavor $\beta$ which can be equal or different of the initial
neutrino flavor $\alpha$.
This is different from the SM CE$\nu$NS process in
Fig.~\ref{fig:CEvNS-Feynman},
where the initial and final neutrino flavors are equal,
because the SM neutral-current weak interaction conserves flavor.
In models BSM there can be flavor-changing
neutral-current interactions.
Unfortunately, the change of neutrino flavor is not observable
in the current and foreseeable experiments,
because the final neutrino cannot be detected
(the only observable signal of CE$\nu$NS is the recoil of the nucleus,
as explained above).
However,
specific flavor-changing NSI can be probed
by their contribution to CE$\nu$NS.

The possible mediators
$\gamma$,
$Z'$, and
$\Phi$
in Fig.~\ref{fig:CEvNS-BSM}
correspond to different scenarios of physics beyond the SM:

I) The photon mediates electromagnetic interactions.
In the SM and in the common belief
neutrinos are exactly neutral and do not interact electromagnetically.
However,
in models BSM
neutrinos can have small electromagnetic interactions~\cite{Giunti:2014ixa}
which contribute to CE$\nu$NS~\cite{Abdullah:2022zue,Giunti:2022aea},
due to three non-standard neutrino electromagnetic properties:
a very small electric charge (often called ``millicharge'')~\cite{Das:2020egb},
a small magnetic moment, and a small anomalous charge radius, which is the only neutrino electromagnetic property that is non-zero already in the SM.
So far there is no experimental indication
in favour of BSM neutrino electromagnetic interactions,
but the search is very active~\cite{Giunti:2022aea}.
Several studies have probed neutrino electromagnetic interactions
through the analysis of CE$\nu$NS data
\cite{Papoulias:2017qdn,Papoulias:2019txv,Miranda:2020tif,Khan:2019cvi,Cadeddu:2018dux,Cadeddu:2019eta,Cadeddu:2020lky,AtzoriCorona:2022qrf,Coloma:2022avw,Liao:2022hno,DeRomeri:2022twg},
finding competitive upper limits. In particular, CE$\nu$NS processes permit to improve the limit on the $\nu_{e}$ charge radius and represent the only existing laboratory bounds on the muonic and tauonic neutrino electric charges~\cite{AtzoriCorona:2022qrf}. Thanks to the lower energy of reactor antineutrinos and the low energy threshold of semiconductor detectors, the data from Dresden-II provide complementary information with respect to CE$\nu$NS processes observed with $\nu$'s produced at SNSs, with negligible dependence on the neutron distribution even if with larger backgrounds and larger dependence on the so-called quenching factor that is not well-constrained at low energies. 

II) The existence of a new massive $Z'$ vector boson
is predicted in several BSM scenarios~\cite{Langacker:2008yv}.
Its mass can be very high,
much larger than the so-called ``electroweak scale'' of about 100 GeV,
or even as low as about 100 MeV.
The search for the possible effects of these bosons is very active in different types of experiments~\cite{Han:2019zkz},
including CE$\nu$NS
\cite{Liao:2017uzy,Papoulias:2017qdn,Coloma:2017ncl,Billard:2018jnl,Papoulias:2019txv,Khan:2019cvi,Giunti:2019xpr,Flores:2020lji,Cadeddu:2020nbr,AtzoriCorona:2022moj,AtzoriCorona:2022qrf,Coloma:2022avw,Liao:2022hno,DeRomeri:2022twg}.  Existing data permit to put limits on a variety of vector boson mediator models:
the so-called universal,
the $B-L$ and other
anomaly-free $U(1)'$ gauge models with direct couplings of the new vector boson with neutrinos and quarks,
and the anomaly-free
$L_e-L_\mu$,
$L_e-L_\tau$, and
$L_\mu-L_\tau$
gauge models where the coupling of the new vector boson with the quarks
is generated by kinetic mixing with the photon at the one-loop level. The constraints related to models where the muon couples to the new boson mediator can also be compared with the values that can explain the muon $g-2$ anomaly~\cite{Muong-2:2021ojo} and further data will be needed to fully exclude these interpretations of the anomaly~\cite{AtzoriCorona:2022moj}.\\

III) The existence of a new massive $\Phi$ scalar boson
is more exotic,
but we cannot miss the chance to probe it in CE$\nu$NS. Interesting constraints have already been put using existing data~\cite{Papoulias:2017qdn,Billard:2018jnl,Papoulias:2019txv,Khan:2019cvi,AtzoriCorona:2022moj,AtzoriCorona:2022qrf,Coloma:2022avw,Liao:2022hno,DeRomeri:2022twg}.\\

Furthermore, even more exotic models as generalized neutrino interactions
\cite{Chang:2020jwl,DeRomeri:2022twg,Majumdar:2022nby},
scalar leptoquarks~\cite{Calabrese:2022mnp},
and extra dimensions~\cite{Khan:2022bcl} can be probed in CE$\nu$NS.

\section{Conclusions}
\label{sec:Conclusions}

The CE$\nu$NS era started with the first measurement of such a process by the COHERENT
experiment in 2017~\cite{COHERENT:2017ipa},
44 years after its theoretical prediction in 1973~\cite{Freedman:1973yd}.
Since then there is great interest in this field,
with many experiments already operating or planned
\cite{Akimov:2022oyb,CONUS:2020skt,CONNIE:2021ggh,MINER:2016igy,NEON:2022hbk,NUCLEUS:2019igx,Akimov:2022xvr,Colas:2021pxr,Baxter:2019mcx,Abdullah:2022zue} that will provide complementary information.
The aim is to observe CE$\nu$NS in different nuclei, confirming the theoretical prediction,
and to obtain from the data information
on the radius of the neutron distribution of the nucleus,
on the weak mixing angle,
and on new neutrino interactions.
The search for effects of physics beyond the Standard Model
is the main goal of current theoretical and experimental research in
particle physics, because we know that the Standard Model,
albeit very successful,
cannot be the final theory of everything:
for example, it cannot explain the smallness of neutrino masses,
the existence of Dark Matter and Dark Energy,
the matter-antimatter asymmetry in the Universe,
and does not include gravity in a consistent theory~\cite{Coloma:2022dng}.
Moreover, CE$\nu$NS is also important for the search for Dark Matter,
because CE$\nu$NS events produced by solar and atmospheric neutrinos
are expected to be detected in future experiments with high sensitivity.
This detection would be very interesting for the study of neutrino properties,
but it constitutes a background for Dark Matter searches
that can limit their sensitivity,
the so-called ``neutrino floor/fog''~\cite{Cooley:2022ufh,OHare:2021utq}, making thus of utmost importance to precisely study CE$\nu$NS in direct detection dedicated experiments.\\
For all these reasons,
we expect interesting CE$\nu$NS news in the next years and we look forward to the future.

\acknowledgments
The work of C. Giunti is supported by the research grant ``The Dark Universe: A Synergic Multimessenger Approach'' number 2017X7X85K under the program ``PRIN 2017'' funded by the Italian Ministero dell'Istruzione, Universit\`a e della Ricerca (MIUR).

\bibliographystyle{h-elsevier}
\bibliography{main}

\end{document}